\documentclass{article}

\usepackage{amsmath}

\usepackage{amssymb}

\usepackage{graphicx}

\usepackage{hyperref}

\usepackage{cite}

\title{ Another look on the connections of Hubble tension with the Heisenberg Uncertainty Principle  }

\author{Oem Trivedi \footnote{oem.t@ahduni.edu.in} \\ School of Arts and Sciences \\ International Centre for Space and Cosmology \\   Ahmedabad University \\  Navrangpura, Ahmedabad, 380009  \\ Gujarat,India}

\begin{document}
	
	\maketitle
	
	\begin{abstract}
		The Hubble tension is one of the most exciting problems that Cosmology faces today. A lot of possible solutions for it have already been proposed in the last few years, with a lot of them using a lot of new and exotic physics ideas to deal with the problem. But it was shown recently that the H0 tension might not require new physics but only a more accurate discussion of measurements and interestingly it was the Heisenberg uncertainty principle which was pivotal for that revelation. Accordingly, if one observed the photon mass beyond the indeterminacy through uncertainty then one could in principle reconcile the tension. We examine this in a greater detail in this work by taking into account modifications of the Compton wavelength, which was pivotal in the initial discussion on the interconnection between uncertainty and the H0 tension. We mainly discuss two types of modifications, one based on generalized uncertainty principles (GUP) and the other on higher dimensional physics considerations. We firstly show that both minimal length and maximal momentum GUP based modifications do not provide photon mass values on the required scales and hence we cannot address the tension in this sense at all in this case. We then show that one can get the photon mass to be on the required scales even after incorporating higher dimensional effects, one cannot reconcile the Hubble Tension in the same sense as in the (3+1) space-time case. In this way, we show that certain new physics considerations cannot address the H0 tension in the same sense as one can using the original Heisenberg uncertainty principle in a (3+1) space-time. 
	\end{abstract}
	
	 \section{Introduction}
	 
	 The advent of precision cosmology has ushered in a new era of explorations of the cosmos. The availability of a plethora of accurate cosmological observations have allowed us to constrain and test various theories to unprecedented levels. This, however, has also sprung up some discrepancies which seem to be pointing towards the limits of our current understanding of the universe. Arguably the most captivating of these new found discrepancies is the Hubble tension which concerns the conflicting measurements of the $H_{0}$ parameter.  Detailed Cosmic Microwave Background Radiation(CMB) maps, combined with Baryon Acoustic Oscillations data, have pointed towards $H_{0} = 66.88 \pm 0.92 $ km $  s^{-1} Mpc^{-1} $ \cite{Planck:2018vyg}. While the increasingly accurate data coming from SNeIa have allowed for a direct estimation of the expansion rate with incredible precision, pointing to a value of $H_{0} = 74.03 \pm $ km $  s^{-1} Mpc^{-1}  $ \cite{riess2019large} with the more recent analysis deriving a value of $H_{0} = 73.04 \pm 1.04$ km $  s^{-1} Mpc^{-1} $ \cite{riess2021comprehensive} and thereby taking the tension to a $ 5 \sigma $ level. 
	 \\
	 \\
	 A huge amount of possible solutions addressing this Tension have been proposed in recent years (see \cite{DiValentino:2021izs} for an extensive discussion on the same) and the central motivations behind most of the solutions are based around considering non-standard cosmological scenarios or more careful discussions of the systematics of the problem \cite{verde2019tensions,vagnozzi2020new,benetti2018h0,graef2019primordial,bernal2016trouble,krishnan2021running,aluri2022observable,Colgain:2022rxy,aghababaei2021hubble,alestas2021late,Aghababaei:2021gxe,Odintsov:2022eqm,Odintsov:2022umu,Dainotti:2021pqg,Dainotti:2022bzg,Yang:2018euj,Pan:2019gop,Yang:2018uae,DiValentino:2019exe,DiValentino:2019ffd,DiValentino:2020naf,DiValentino:2019jae,Nunes:2021zzi,Anchordoqui:2021gji,Alestas:2021luu,Poulin:2018cxd,Smith:2020rxx,Poulin:2021bjr,Smith:2022hwi,Kreisch:2019yzn,Cyr-Racine:2021oal,Knox:2019rjx,Vagnozzi:2021tjv,Reeves:2022aoi,Vagnozzi:2021gjh,Agrawal:2019lmo}. So a key point of debate still remains about whether we need to use new physics to address this issue or instead more carefully go by the required data analysis. On this note a new way looking at the tension,proposed in \cite{capozziello2020addressing,spallicci2021heisenberg}. It was shown in the paper that a measurement of the photon mass \cite{retino2016solar,hagiwara2002review,williams1971new} beyond the indeterminacy limit imposed by the Heisenberg uncertainty principle can reconcile the Hubble tension at z=1 at the 4.4 $\sigma$ level (although the analysis could also be carried over to the most recent SNeIa results as well \cite{riess2021comprehensive}). The key to this result lied in the estimates for the photon mass for both sets of the $H_{0} $ measurements, where $m_{HST} \sim 1.61 \times 10^{-69} kg $ and $ m_{CMB} = 1.46 \times 10^{-69} kg $ (where $m_{HST}$ and $m_{CMB} $ correspond to the photon masses for $H_{0} \sim 74 Km/s/Mpc $ and $H_{0} \sim 67 Km/s/Mpc $ at z=1, respectively). This calculation involved the consideration that the Compton wavelength of the observed universe is given by the luminosity distance \footnote{Such a consideration is very ad hoc, as it is not necessarily straightforward to assume such an equivalence between the luminosity distance and the compton wavelength. One may argue that all distances are connected by powers of (1+z) however it is not generic, though, because this relies on the assumption of the Etherington distance-duality connection. The reason we will still be considering this equivalence in this paper is because we would like to see whether an analysis with similar grounding principles as the one performed in the conventional cosmological settings would hold even for the most basic quantum gravitatioanlly motivated corrections.}. Hence in this sense, addressing the Hubble tension would then only need a more thorough discussion of measurements rather than new physics and while the paper made a nice case for this point, there are some subtle issues with this work. One thing is that the work assumes intrinsically that both the early universe and local values of $H_{0}$ are both correct even though they don't agree with each other, which does not stand right with the standard model of cosmology as it would need two different values of $H_{0}$ for the same cosmology. It also does not has any comments on how local measurements of $H_{0}$ all go higher than 70. Finally, the proposition of measuring the photon mass to the levels at which the results of this analysis would bear fruits is something which is practically impossible to expect. Hence even though the study seems appealing in principle, one would not be very optimistic of actually alleviating the issues of the Hubble tension using this. However, it would nevertheless be very interesting to actually see what similar analysis would say about the $H_{o}$ tension considering some kinds of exotic physics. So in this work we discuss these issues in detail and look to see whether one can have the same kind of assertions as in \cite{capozziello2020addressing} for $H_{0} $ tension if one starts to consider some new physics in this scenario. Particularly, we will consider 2 types of modifications of the Compton wavelength formula where one is based on Generalized Uncertainty Principles \cite{Maggiore:1993rv,Adler:2001vs,Tawfik:2015rva,Scardigli:1999jh,Scardigli:2003kr,Scardigli:2010gm,Scardigli:2014qka,Vagnozzi:2022moj,Okcu:2021oke,Jusufi:2020wmp,Neves:2019lio,Scardigli:2014qka} while the other one is based on higher dimensional physics considerations. Generalized uncertainty principles are generalizations of the Heisenberg Uncertainty principle which are based on considerations from various quantum gravity approaches like String theory, Loop quantum gravity etc and also from various topics of Black hole physics.  One can write the Compton formula for a GUP with minimal length considerations as \begin{equation}
	 \lambda_{c} = \frac{\hbar}{mc} \left(1 + \alpha \frac{m^2}{m_{p}^2}\right)
	 \end{equation} where $m_{p} $ is the planck mass. For a GUP with both minimal length and maximal momentum considerations as \begin{equation}
	 \lambda_{c} = \frac{\hbar}{mc ( 1 - \beta m^2 c^2 )}
	 \end{equation} One can also write the Compton formula for a space time with (3+n) spatial dimensions as \cite{lake2016compton} \begin{equation}
	 \lambda_{c} = \left( \frac{\hbar}{m c} \right)^\frac{1}{1+n} (R_{e})^\frac{n}{1+n} 
	 \end{equation} Where $R_{e} $ is assumed to be a single length scale at which all the extra dimensions in which matter is free to propagate is compactified and n is the number of extra spatial dimensions (for example, for n=1 we would have 4 spatial dimensions etc.). In the next section we will consider the effects of the modified compton formulas from the GUP based considerations on the Hubble tension while in section III we consider modified Compton formula from higher dimensional considerations on the same and we finally conclude our discussion in Section IV.
	 \\
	 \\
	 \section{Incorporating GUP based effects }
	 Modifications of the Compton formula from GUP based considerations have gained attention recently (see \cite{Lake:2015pma}). The existence of a minimal length and a maximum momentum accuracy is preferred by various physical observations. Furthermore, assuming modified dispersion relation allows for a wide range of applications in estimating, for example, the inflationary parameters, Lorentz invariance violation, black hole thermodynamics, Saleker-Wigner inequalities, entropic nature of the gravitational laws, Friedmann equations, minimal time measurement and thermodynamics of the high-energy collisions. One of the higher-order GUP approaches gives predictions for the minimal length uncertainty. Another one predicts a maximum momentum and a minimal length uncertainty, simultaneously. For the modification as described in (1), The GUP one needs to consider is of the form \begin{equation}
	 \Delta x \geq \frac{\hbar}{\Delta p} + \frac{\alpha l_{p}^2 \Delta p}{\hbar}
	 \end{equation} We see that the modified compton formula is obtained when one considers $ \Delta x \to \lambda $ and $\Delta p \to mc $. The above GUP is the generalization of the uncertainty principle which about from minimal length considerations and is known as the Kempf-Mangano-Mann (KMM) GUP. The KMM GUP can be deduced from following commutator \begin{equation}
	 [\hat{x},\hat{p}] = i \hbar(1 \pm \alpha \hat{p}^{2})
	 \end{equation}  
	 where $\alpha$ can be either positive or negative. Its's important to stress though that the magnitude of $\alpha$ has been widely studied in the literature and has been seen to be positive mostly. For instance, in String theory it has been assumed to be of order unity \cite{amati1987superstring,amati1989can,maggiore1994quantum}. From a straightforward analysis of the quantum corrections to the Newtonian potential, one can derive $\alpha = \frac{82 \pi}{5} $ \cite{Scardigli:2016pjs} whereas by studying the deformed Unruh temperature in a maximal acceleration framework one would get $\alpha = \frac{8 \pi^2 }{9} $. Similar results have been found in several wide ranging contexts in things like non-commutative Schwarzschild geometry \cite{Kanazawa:2019llj} and even in the Corpuscular description of Black Holes \cite{Buoninfante:2019fwr}. While the KMM GUP is based on minimal length considerations, the GUP on which (2) is based on has the form \begin{equation}
	 \Delta x \geq \frac{1}{\Delta p (1 - \beta p^2)}
	 \end{equation}
	 where $\beta$ is again a parameter which can take on both positive and negative values. This GUP is due to Pedram \cite{pedram2012higher,pedram2012higher2} and has both minimal length and maximal momentum considerations. Again, the magnitude of $\beta$ has been seen to not be of any extreme magnitudes in the literature. 
	 \\
	 Now that we are done with this, we start exploring the effects of these modifications on the Hubble tension. For a more complete description here which can also be consistent with higher redshifts, the usual Hubble law does not work and we need to work with appropriate cosmographic series or even other more detailed approximants like the Pade Series \cite{aviles2014precision}. For our purposes, we can simply adopt
	 a cosmographic Taylor series truncated at the second order where kinematics
	 and dynamics of Hubble-Lemaitre flow have to be considered. The luminosity distance $d_{L} (z) $ can be written as \begin{equation}
	 d_{L} (z) = \frac{zc}{H_{0}} \left(1 + \frac{z}{2} (1-q_{o})\right)
	 \end{equation}
	 where $q_{o} $ is the cosmographic deceleration parameter of the universe. The parameter can be constrained to be $ -0.64 \pm 0.22 $ using
	 the joint Pantheon data for SNeIa with BAO and time-delay measurements by
	 H0LiCOW and angular diameter distances measured using water megamasers \cite{capozziello2019model} or $q_{o} = -0.28 \pm 0.49 $ using JLA compilation SNeIa data with BAO and observational Hubble parameter values \cite{capozziello2019extended}. Now we equate the luminosity distance to the Compton wavelength \cite{capozziello2020addressing} for the KMM GUP (1) as \begin{equation}
	 \frac{\hbar}{mc} \left(1 + \alpha \frac{m^2}{m_{p}^2}\right) = \frac{z c}{H_{0}} \left( z + \frac{1}{2} (1-q_{o}) \right) 
	 \end{equation}We can choose z = 1 and $q_{o} = -1/2 $ for the sake of simplicity here to get \begin{equation}
	 \frac{\hbar}{mc} \left(1 + \alpha \frac{m^2}{m_{p}^2}\right) = \frac{7c}{4 H_{0}}
	 \end{equation} From which we get the equation \begin{equation}
	 \alpha \frac{m^{2}}{m_{p}^2} - \frac{7cm}{4 H_{0}} + \frac{\hbar}{c} = 0
	 \end{equation} If we ignore the term $\frac{\hbar}{c}$ for how small it is as compared to the other terms, one can now arrive at the mass of the photon in this case to be \begin{equation}
	 m = \frac{7 c^2 m_{p}^2}{4 \alpha \hbar H_{0}} \approx \frac{10^{53}}{\alpha}
	 \end{equation} As is clear from above, the predicted mass scale for the photon is incredibly large in this case and for it to be even $\mathcal{O}(1) $, one needs a radically large value of $\alpha$ which is of the order of the $10^{53}$ and such values would not be appreciated and would really hamper the formulation of the GUP itself, besides no such values have been encountered in the considerations which have led to the formulation of the GUP itself. The photon mass has been constrained to $m_{\gamma} < 10^{-54} $ kg by the Particle Data Group from measurements in the solar system \cite{hagiwara2002review} and from laboratory tests the upper bound is instead found to be $m_{\gamma} < 10^{-50} $ kg \cite{williams1971new}. Hence, getting $\alpha$ to these levels would require it to be comfortably greater than $\mathcal{O} (10^100) $. One thing to note is that the equation (4) is properly a quadratic equation and we neglected the contribution of the term $ \frac{\hbar}{c} $, but similar observations for $\alpha$ would remain even if we solved the equation properly using the quadratic formula without this approximation.
	 \\
	 \\
	 Furthermore, one can also talk about other kinds of GUP's besides the KMM formulation. We can for example talk about the Pedram GUP \cite{Pedram:2012my} , which can be written like the KMM GUP as \begin{equation}
	 \Delta x \geq \frac{1}{\Delta p (1 - \beta p^2)}
	 \end{equation} We can again get the modified Compton formula for this GUP in a similar to way to KMM GUP by making the considerations $ \Delta x \to \lambda $ , $ \Delta p \to mc $. The modified compton formula becomes \begin{equation}
	 \lambda = \frac{\hbar}{mc ( 1 - \beta m^2 c^2 )}
	 \end{equation} Again making the same considerations on the luminosity distance, we arrive a the formula for the photon mass in this case to be given by the cubic equation  \begin{equation}
	 m^3 \beta c^2 - m + \frac{4 H_{0} \hbar}{7 c^2} = 0 
	 \end{equation} If we choose to neglect the term $\frac{4 H_{0} \hbar}{7 c^2}$ for how small it is, we get that \begin{equation}
	 m \approx \frac{10^{-8}}{\sqrt{\beta}}
	 \end{equation} We see that it would need very large values of $\beta$ to make the predicted mass scale here to be anywhere close to the nearest estimates of around $ 10^{-50} $ by laboratory tests, which would need $ \beta \sim \mathcal{O} (10^{84}) $. Again, just like the KMM GUP approach, the values of these parameters being this high would seriously hamper their own formulations and considerations and one would not expect such high values for these parameters. One thing we note again is that if we did not drop out the term $\frac{4 H_{0} \hbar}{7 c^2}$ in (8) and still consider it, the results for the mass scale would not change very much (and so even changing the $H_{0} $ values between 67 and 74 would not matter much and we would not have $ \frac{\Delta m}{m} \sim 0.1 $ in any case). Changing the values of $q_{o} $ here would not change the main results obtained in this analysis. The principle inference we can draw here is that GUP's cannot point towards the same conclusions that the Heisenberg Uncertainty principle in its original form could, with regards to the Hubble tension \footnote{One can also talk about GUPs where non-linear corrections apply to both position and momentum like \cite{Bambi:2007ty} however making the same considerations as for the cases we have discussed here with $\Delta x \to \lambda $ and $\Delta p \to mc $, one can recover similar conclusions but we will not be discussing it in detail here. }. One can only explain the Hubble tension with the uncertainty principle if one considers the original Heisenberg uncertainty principle, not the ones coming from phenomenological considerations based in string theory, Loop quantum gravity. etc.
	 \\
	 \\
	 \section{Incorporating higher dimensional effects}
	 Modifications of the Compton formula owing to higher dimensional physics considerations were made by Lake and Carr in \cite{lake2016compton} where they pursued the so called " Compton-Schwarzschild correspondence ". The correspondence was motivated by the observation that the Compton wavelength and the Scwharzschild radius are dual to each other under the transformation $ m \to \frac{m_{p}^{2}}{m} $. This can be interpreted to be suggesting a link between elementary particles in the $ m < m_{p} $ regime and Black holes in the $ m > m_{p} $ regime. In the presence of n extra spatial one may expect this duality to break however it was shown that it can be restored as the effective Compton wavelength depends on the form of a (3+n) dimensional wavefunction and if it is spherically symmetric, the duality is restored. The higher dimensional Compton wavelength formula can be written as in (3) with the length scale having the following form for different values of n \cite{arkani1998hierarchy}
	 \begin{equation}
	 R_{e} \sim 10^{(32/n) - 17} cm
	 \end{equation}
	 Now we can again start to incorporate the effects of these modifications for the Hubble tension. We begin again by assuming that the Compton wavelength is equal to the luminosity distance and so \begin{equation}
	 \left( \frac{\hbar}{m c} \right)^\frac{1}{1+n} (R_{e})^\frac{n}{1+n} = \frac{z c}{H_{0}} \left( z + \frac{1}{2} (1-q_{o}) \right) 
	 \end{equation}   
	 As done before, we set z = 1 and $q_{o} = -1/2 $ to get \begin{equation}
	 \left( \frac{\hbar}{m c} \right)^\frac{1}{1+n} (R_{e})^\frac{n}{1+n} = \frac{7c}{4 H_{0}}
	 \end{equation}
	 And so m can be written as \begin{equation}
	 \frac{\hbar}{mc} R_{e}^n = \left( \frac{7c}{4 H_{0}} \right)^{(1+n)} \to \frac{\hbar R_{e}^n}{c} \left( \frac{4 H_{0}}{7c} \right)^{(1+n)} = m
	 \end{equation}
	 Now, we compute m for HST by setting $H_{0} \sim 73$ km/s/Mpc and for a (4+1) Space-time (n=1, $R_{e} = 10^10 $ km) and we get, \begin{equation}
	 m_{HST} = 7.13 \times 10^{-89} kg
	 \end{equation}
	 While for CMB ($H_{0} \sim 68 $ km/s/Mpc), we have \begin{equation}
	 m_{CMB} \approx 6.00 \times 10^{-89} kg
	 \end{equation}
	 Now even though the photon mass scale values are within the requirements set by the Particle data group and Laboratory tests \cite{hagiwara2002review,williams1971new}, we now have \begin{equation}
	 \frac{\Delta m}{m} \sim 0.2
	 \end{equation}
	 which is a bit higher than $ \frac{\Delta H_{0}}{H_{0}} \sim 0.1 $ . If we now perform the same calculations for n = 2 or n=7 (String/M theory scale) or for an arbitrarily large number of spatial dimensions $n \to \infty $ , the predicted mass scale for the photon will keep on going down (comfortably reaching below $\mathcal{O} (10^{-100})$) while $ \frac{\Delta m}{m} $ would keep on increasing further and further, thereby again not being able to explain the Hubble tension quite in the same way as in \cite{capozziello2020addressing}. In fact, it is easy to check that the value of $ \frac{\Delta m}{m} $ keeps on increasing as we increase the number of extra spatial dimensions up until the String/M-Theory levels, where $ \frac{\Delta m}{m} \sim 1 $. This makes it clear that the extra dimensional considerations cannot provide the same inferences for the Hubble tension as one gets in the usual (3+1) space-time case \cite{capozziello2020addressing}. 
	 \\
	 \section{Concluding remarks }
	 In this work we have again explored possible links of the Hubble tension with the Uncertainty principle in greater varieties. The $H_{0}$ tension presents a very exciting scenario for cosmology and while a lot of possible solutions have been proposed, it becomes important to take a pause and consider whether we really need new physics to solve the problem or a better understanding of the data measurements itself. Previous work showed interconnections of the Hubble tension and the Heisenberg Uncertainty principle and how the tension can be reconciled if one observes the mass of the photon to be beyond the Heisenberg indeterminacy limit. Thereby, it would appear that we do not need new physics to address the $H_{0}$ tension but rather a better gauge of the measurements themselves. To probe this assertion in greater detail, we considered two "new physics" based modifications of the Compton wavelength. The first of these were the Generalized uncertainty principles, which are generalizations of the Heisenberg Uncertainty based on considerations mainly from quantum gravity approaches and black hole physics. We showed that one cannot recover viable mass scales for the photon in this case, if one goes by the initial assumption of the equality of the Compton wavelength and Luminosity distance. Thereby explaining the Hubble tension in the same way as done for the unmodified Compton formula (or equivalently, by considering only the original Heisenberg uncertainty principle) is not possible. The second new physics based modification of the Compton formula that we worked on came by due to higher dimensional physics considerations(mainly based on black hole physics), through the "Compton-Schwarzschild correspondence". We showed that although in this case one can recover the photon mass to be within the viable limits as imposed by the experiments (although very low, much lower than what the foreseeable sensitivity of the experiments would be) \cite{hagiwara2002review,williams1971new}, one can still not really explain the Hubble tension at the 5 $\sigma$ in the same way as in the usual (3+1) space-time case. To conclude we have shown that one cannot reconcile the $H_{0}$ tension with measurement based considerations if one considers modifications through quantum gravity phenomenology and black hole physics. This goes to show that the links of the $H_{0}$ tension with the original Heisenberg uncertainty principle as shown in \cite{capozziello2020addressing} can only be held true in the most conventional of cosmological settings and really goes to show that new physics based considerations rule out any such connections in the grand scheme of things.
	 
	 \section*{Acknowledgements}
	 
	 The author would like to thank Salvatore Capozziello, Alessandro Spallicci and Micol Boneti for their comments on the paper. The author would also like to thank the reviewer for their insightful comments on the manuscripts.

\bibliography{JSPJMJhubble.bib}

\bibliographystyle{unsrt}
	
\end{document}